\documentclass[twocolumn,aps,pra,groupedaddress,showpacs,floatfix]{revtex4}
\usepackage{epsfig,amssymb,amsmath}
\begin{document}

\title{Information-capacity description of spin-chain correlations}

\author{Vittorio Giovannetti and Rosario Fazio}
\affiliation{NEST-INFM \& Scuola Normale Superiore,
piazza dei
Cavalieri 7, I-56126 Pisa, Italy.}
%\date{\today}

\begin{abstract}
Information capacities
%modifiche
achievable in the 
multi-parallel-use scenarios
%modifiche
are employed
to characterize the quantum correlations 
in unmodulated spin chains. By studying the qubit amplitude
damping channel, we calculate the quantum capacity $Q$, 
the entanglement assisted capacity $C_E$, and the classical 
capacity $C_1$ of a spin chain with ferromagnetic
Heisenberg interactions.
\end{abstract}
\pacs{03.67.Hk,05.50.+q,03.67.-a,03.65.Db} 
\maketitle 

\section{Introduction}
Spin chains are gaining increasing attention as natural candidates of quantum 
channels~\cite{BOSE,DATTA,OSBORNE,SETH,BURGARTH}. An unknown quantum state can 
be prepared on one end of the chain and then transferred to the other end   
by simply employing the ability of the chain to propagate the state by means 
of its dynamical evolution. This procedure does not require any gating and 
therefore can be implemented without any need for a  modulation of the 
couplings between the spins. Especially this last aspect may be of importance for
solid state quantum information. Several aspects of quantum communication 
using spin chains have been already obtained starting from 
the original proposal by Bose~\cite{BOSE}. 
The fidelity of the transmitted state 
can be increased if the sender and the 
receiver can encode/decode the state using a finite number of spins~\cite{OSBORNE}.
Even perfect transmission can be achieved if the exchange couplings of the chain 
are chosen appropriately~\cite{DATTA} or if it possible to perform measurements 
on the quantum spins of the chains~\cite{CIRAC}. 
Moreover, it was recently shown that a phase covariant cloner can be 
realized using an unmodulated spin network~\cite{DECHIARA} and that the
spins chain approach to quantum communication 
seems to be realizable in solid state devices with present day technology~\cite{ROMITO}.

Another interesting aspect which is emerging recently is the interconnection
between  quantum information and  condensed matter theory~\cite{PRESKILL}. 
Examples are the study of non-local (quantum) correlations of spin systems 
in a variety of situations (see~\cite{ENT} and references therein) or 
reformulation of the density matrix 
renormalization group in the framework of 
quantum information theory~\cite{DMRG}.
In this paper we would like to further explore the use of concepts born in 
quantum information for the characterization of spin chains. 
Here we propose to use the information capacities 
of  quantum channels as a 
tool to characterize some aspects of dynamical  
correlations in a spin chain. 
By calculating the capacities of the channel, 
obtained by identifying two separate 
sections of the chain as the extremes of a 
communication line, one can in fact get
some information about the strength of the correlations among the 
interconnecting spins.
In some sense this is equivalent to assigning to any couple of subsets 
of the spins in the sample  the values of the corresponding capacities 
in order to create a ``road map'' of the information fluxes in the system.

Our results apply for the whole class of Hamiltonians 
for which the total magnetization along a fixed direction is conserved. 
In the present work 
we confine ourself to the case in which 
one bit is transferred through the chain. 
This situation corresponds to study the sectors 
in which only few spins are up. 
We believe, however that the present approach can be 
further extended to other spin sectors.

The paper is organized in two distinct parts. 
In the first one we introduce the 
spin chain communication lines and we discuss  some
models in details (Sec. \ref{s:uno}). 
In the second part  (Sec. \ref{s:cap}) instead 
we give a brief overview of channel capacities and,
by studying the qubit amplitude damping channel~\cite{CHUANG}, 
we calculate the quantum capacity $Q$, the entanglement
assisted capacity $C_E$ and the classical capacity $C_1$ of
the spin chain models presented in Sec. \ref{s:uno}.
As explained in detail throughout the paper the above quantities
are computed in communication scenarios where 
many ``copies'' of the spin system are available  
and ``not'' for those scenarios where the communicating parties
 make multiple uses of the ``same'' spin chain.
The paper ends in Sec. \ref{s:tre} with the conclusions.

\section{The model}\label{s:uno}
Given  a collection of $N$ spins coupled by means of a time-independent 
Hamiltonian it is possible to define a quantum channel by identifying
two set of spins of the sample (say, sets $A$ and $B$) as two quantum 
registers.
A first party (the sender of the message) encodes some information on $A$ and 
a second party (the receiver) tries to recover such information from $B$ some 
later time $t$~\cite{BOSE,SETH,OSBORNE}. 
Formally, one assumes that the encoding 
procedure takes place by initially decoupling the spins $A$ from the remaining 
spins without disturbing them from their initial fiduciary 
state $\sigma_0$, preparing $A$ on some input message $\rho_A$, 
and finally allowing $\rho_A$ and   $\sigma_0$
to interact for a given time $t$ 
through the Hamiltonian of the system. 
At this point we consider the state $\rho_B(t)$ of 
the spins $B$, obtained from the spin chain state 
$R(t)=U(t) (\rho_A \otimes \sigma_0) 
U^{\dag}(t)$ by tracing away all the the degrees of freedom of the 
system but those
relative to $B$. [Here $U(t)$ is the unitary evolution of the system].
The resulting Completely Positive, Trace preserving 
(CPT) mapping 
\begin{eqnarray}
\rho_A  
\rightarrow { \cal M}(\rho_A )
\equiv \rho_B(t)= 
\mbox{Tr}^{(B)} [ U(t) (\rho_A \otimes \sigma_0)  
U^{\dag}(t)]
\label{mappa}
\end{eqnarray}
where $\mbox{Tr}^{(B)}$ means trace over all the  spins but $B$, 
defines the quantum channel we are interested in. 
A caveat is in order. Equation~(\ref{mappa}) does not provide a proper
description of any realistic scenario where the communicating parties
keep on operating on their quantum registers over an extended period
of time. In fact, after  a first ``reading out'' of the signal 
from the spin chain at time $t$, 
the state $\sigma_0$ of the spins not belonging to $A$ will change, thus making 
Eq.~(\ref{mappa}) unable to describe the state of $B$ 
for later times.
As a matter of fact, in our system
any repeated (in time) manipulation of the quantum registers 
will necessarily introduce memory effects in the communication
for which a satisfying quantum information theory is still missing
(some preliminary results on quantum channels with memory 
can be found in Ref.~\cite{MEMO}).
 
Having the previous observations in mind, Eq.~(\ref{mappa})
can still be used to provide a quite complete characterization of the correlations
between the registers $A$ and $B$.
Technically the map $\cal M$ is defined 
in the Hilbert space $\cal H_A$ of the configurations of $A$ and 
depends on $t$, $H_{int}$, $\sigma_0$ and on the choice of $A$ and $B$.
A common approach \cite{BOSE,OSBORNE,DECHIARA} used to quantify the
correlations between the sets $A$ and $B$ of the chain 
is then to consider the  fidelity 
associated with $\cal M$, by computing the average fidelity  between the input states 
$\rho_A$ and their output counterparts $\rho_B(t)$.
Here we observe that a more detailed characterization of such
correlations can be obtained by treating $\cal M$ of Eq.~(\ref{mappa}) as the CPT 
map of a real memoryless channel. The 
correlations between $A$ and $B$ can then be 
analyzed by means of the  information capacities 
\cite{SHOR,CHUANG,SETHQ,HSW,BENNETT1,BEN} 
associated with~$\cal M$. These quantities are related to the minimum 
``amount of redundancy'' over multiple uses of the channel $\cal M$ needed 
to achieve perfect message transmission, i.e. unitary fidelity: 
higher values of the capacities correspond to stronger correlations 
between input and output states. 
Depending 
on the nature (e.g. classical or quantum) of the information 
propagating through the channel one can define different capacities of $\cal M$
and each of them are obtained by  maximizing over all possible 
coding and decoding strategies which act over multiple channel uses 
(see Sec.~\ref{s:cap}).
From the discussion that follows Eq.~(\ref{mappa}) it should be
evident that the capacities of the map $\cal M$ might not provide a 
proper description of the communication performances of the spin chain. 
As a matter of fact, they only account for those scenarios where
many parallel ``copies'' of the same system are simultaneously operated  by the two
communicating parties but fail to describe those scenarios where the communicating parties
 make successive multiple uses of the ``same'' spin chain.

\subsection{Solvable models}\label{s:solvable}
Examples of a spin chain channel where the information capacities 
of $\cal M$ can
be solved exactly are provided by a  chain of $1/2$-spins coupled 
through a ferromagnetic Heisenberg interaction~\cite{BOSE}.
The system  Hamiltonian is
\begin{eqnarray} 
H=-\sum_{\langle i,j \rangle} \hbar 
J_{ij} \left({\sigma}_x^{i}{\sigma}_x^{j}
+{\sigma}_y^{i}{\sigma}_y^{j}+\gamma {\sigma}_z^{i}{\sigma}_z^{j}
\right)
-\sum_{i=1}^{N} \hbar B_i \sigma_z^{i} \;,
\label{hamilto}
\end{eqnarray}
where  the first summation is performed on the nearest-neighbor spins of
the chain, $\sigma^i_{x,y,z}$ are the Pauli operators associated with the
$i$th spin, $J_{ij}$ are coupling constants, $\gamma$ is an anisotropy 
parameter, and $B_i$ are associated with externally applied magnetic fields.

In the following we will assume  the input set $A$ and output set $B$ 
of Eq.~(\ref{mappa}) to 
contain, respectively, the first $k$ and the last $k$ spins of the chain, 
and that all the spins but $A$ are initially prepared in the same 
eigenstate $|\downarrow \;\rangle$ of $\sigma_z$. 
[Here $|\downarrow \; \rangle$ 
and $|\uparrow \;\rangle$ are the eigenstates of $\sigma_z$ associated 
respectively with the eigenvalues $-1$ and $+1$].
If the two communicating parties 
are allowed to use the entire Hilbert space of their quantum memory, 
Eq.~(\ref{mappa}) yields a $k$-qubit channel which,
in general, is too complex to be  analyzed on a complete basis. 
In the present paper
we consider thus a simplification of this scenario
where the sender and the receiver use their $k$ spins to encode
a single logical qubit.
This approach is clearly inefficient from an 
informational theoretical point of view,
but on the positive side, the resulting maps can be treated analytically.

\subsubsection{Encoding one logical qubit with one-spin up
vectors}\label{primavariazione}

As in Refs.~\cite{BOSE,OSBORNE} we introduce the one-spin up vector
\begin{eqnarray}
| { j}\rangle \equiv |\downarrow \downarrow 
 \cdots \downarrow \uparrow \downarrow \cdots 
\downarrow \rangle
\;, \label{vettori}
\end{eqnarray}
which for $j=1,\cdots,N$ represents the state of the chain where the $j$th spin 
is prepared in the eigenstate $|\uparrow \; \rangle$ and the other $N-1$ ones
in $|\downarrow\;\rangle$. 
Suppose that at time $t=0$ the sender prepares her/his spins in
\begin{eqnarray}
|\Psi\rangle_A \equiv \alpha |\Downarrow\;\rangle_A + 
\beta|\phi_1 \rangle_A \;, \label{inputA1}
\end{eqnarray}
where $\alpha$, $\beta$ are complex amplitudes, 
$|\Downarrow \; \rangle_A$ is the state of $A$ with all spins down, and
$|\phi_1\rangle_A$ 
is a given normalized superposition of $| { j}\rangle$ with
$j$ referring to spins of the input memory $A$, i.e.
\begin{eqnarray}
|\phi_1\rangle_A \equiv \sum_{j=1}^k c_j |{ j} \rangle \;.
\label{PHI1}
\end{eqnarray}
Since the Hamiltonian of Eq.~(\ref{hamilto}) commutes 
with the total spin component along the $z$ direction, one can show 
that at time $t$ the whole chain is described by \cite{OSBORNE}
\begin{eqnarray}
|\Psi(t)\rangle \equiv \alpha |\Downarrow \; \rangle + \beta \sum_{j^{\prime}=1}^{N}
\sum_{j=1}^k c_{j}f_{j^{\prime},j}(t) |{ j^{\prime}}
\rangle \;,
\label{OUTPUT1}
\end{eqnarray}
with $|\Downarrow\;\rangle$ the state of the chain with all spins down and with 
\begin{eqnarray}
f_{j,s}(t)\equiv
\langle { j} | e^{-i 
H t/\hbar}|
{ s}\rangle \label{effe}\;
\end{eqnarray}
[see, for instance, Ref.~\cite{BOSE} 
for the explicit functional dependence of $f_{j,s}(t)$ from the
evolution time $t$].
According to Eq.~(\ref{mappa}) 
the state of $B$ at time $t$ is finally obtained from  Eq.~(\ref{OUTPUT1})
by tracing over all the remaining $N-k$ spins, i.e.
\begin{eqnarray}
\rho_B(t) &=& (|\alpha|^2 + (1-\eta) |\beta|^2) | \Downarrow \rangle_B\langle 
\Downarrow | + \eta |\beta|^2 |\phi_1^{\prime}\rangle_B \langle \phi_1^{\prime}|
 \nonumber \\
&+& \sqrt{\eta} \alpha \beta^* | \Downarrow \rangle_B\langle 
\phi_1^{\prime} | +  \sqrt{\eta} \alpha^* \beta | \phi_1^{\prime} \rangle_B\langle 
\Downarrow |
\label{rhodibeta1}\;,
\end{eqnarray}
with
\begin{eqnarray}
\eta = \sum_{j^{\prime}=N-k+1}^N \left|\sum_{j=1}^k c_j f_{j^{\prime},j}(t)\right|^2 
\label{etanew}\;.
\end{eqnarray}
In Eq.~(\ref{rhodibeta1}) the two orthonormal vectors  
$|\Downarrow \;\rangle_B$ and $|\phi_1^{\prime}\rangle_B$ 
are, respectively, the state of $B$ with all spin down, and 
\begin{eqnarray}
|\phi_1^{\prime}\rangle_B \equiv \sum_{j^{\prime}=N-k+1}^{N} 
\sum_{j=1}^k c_j f_{j^{\prime},j}(t)  |{ j^{\prime}} 
\rangle /\sqrt{\eta} \;.
\label{PHIPRIME1}
\end{eqnarray}
Apart from an irrelevant unitary transformation \cite{NOTABENE}
the map associated with $\rho_B(t)$ of~(\ref{rhodibeta1}) 
is a qubit amplitude damping channel \cite{BOSE,OSBORNE} of efficiency $\eta$ 
which acts 
on the orthonormal basis $\{ |\Downarrow \;\rangle_B,
|\phi_1^{\prime}\rangle_B \}$.
For the sake of clarity let us
identify $| \Downarrow \;\rangle_A$ and $| \Downarrow \;\rangle_B$ with
the same logical qubit state $|0\rangle$ and $| \phi_1\rangle_A$,
$| \phi_1^\prime \rangle_B$ with $|1\rangle$. In this notation we can
express the input state $|\Psi\rangle_A$ and the output state $\rho_B(t)$
as density matrices $\rho$ and $\rho^\prime$ of the same qubit Hilbert 
space ${\cal H}_A$. 
Equation~(\ref{rhodibeta1}) becomes thus
\begin{eqnarray}
\rho^\prime = {\cal D}_{\eta} (\rho)\;,
\label{rhob}
\end{eqnarray}
with ${\cal D}_\eta$ the amplitude damping map characterized by the 
Kraus operators \cite{CHUANG}
\begin{eqnarray}
A_0&=& |0\rangle\langle 0| +\sqrt{\eta}\;|1\rangle \langle 1| \;, \nonumber \\
A_1&=& \sqrt{1-\eta}\;|0\rangle \langle 1| \;.\label{kraus}
\end{eqnarray}
Equation~(\ref{rhob}) describes a quantum channel in which
the logical information of $A$ represented by the
coefficients $\alpha$ and $\beta$ of Eq.~(\ref{inputA1}), is transferred to 
the output memory $B$ with an accuracy which can be estimated by
calculating the capacities of the map ${\cal D}_\eta$. 
The calculation of the capacities
of ${\cal D}_\eta$ is presented in Sec.~\ref{s:thecapa}.

\subsubsection{Encoding one logical qubit with 
two-spin up vectors}\label{secondavariazione}

Consider the case where in Eq.~(\ref{inputA1})
the vector $|\phi_1\rangle_A$  is replaced by
a normalized superposition $|\phi_2\rangle_A$ of states 
$|{ j},
{ \ell} \rangle$  where the $j$-th and $\ell$-th 
spins  of the chain are in $| \uparrow\;\rangle$ while the remaining are 
in $|\downarrow \;\rangle$, i.e.
\begin{eqnarray}
|\phi_2\rangle_A \equiv \sum_{j>\ell=1}^k d_{j,\ell} |{ j},
{ \ell} 
\rangle \;.
\label{PHI2}
\end{eqnarray}
In other words, the sender still uses the
state with no spin up to ``transfer'' $\alpha$,
but now a selected superposition of two-spin up states 
is employed to ``transfer'' $\beta$.
In this case one can show that the output state of the memory $B$ is
\begin{eqnarray}
\rho_B(t) &=& (|\alpha|^2 + \eta_2 |\beta|^2) | \Downarrow \rangle_B\langle 
\Downarrow | + \eta_1 |\beta|^2 |\phi_2^{\prime}\rangle_B \langle \phi_2^{\prime}|
 \nonumber \\
&+& \sqrt{\eta_1} \alpha \beta^* | \Downarrow \rangle_B\langle 
\phi_2^{\prime} | +  \sqrt{\eta_1} \alpha^* \beta | \phi_2^{\prime} \rangle_B\langle 
\Downarrow | \nonumber \\
&+& (1-\eta_1 -\eta_2) |\beta|^2 \sigma_B
\label{rhodibeta2}\;,
\end{eqnarray}
where $|\phi_2^{\prime}\rangle_B$ is a superposition of two-spins up state of
$B$ and $\sigma_B$ is a density matrix of one-spin up states of $B$ whose 
eigenvectors are orthogonal with respect to  \mbox{$| \Downarrow \rangle_B$} and 
$|\phi_2^{\prime}\rangle_B$
(see Appendix~\ref{a:postpo} for details). For $\eta_1+\eta_2= 1$ 
the mapping (\ref{rhodibeta2}) reduces to 
a qubit amplitude damping channel ${\cal D}_\eta$ with quantum efficiency 
$\eta=\eta_1$.
However, in the general case, 
Eq.~(\ref{rhodibeta2}) is slightly more complex. In fact for $\eta_1+\eta_2<1$
the component $|\phi_2\rangle_A$ of the input 
state undergoes to three possible
processes:  with probability $\eta_1$ 
it is rotated into the output state $|\phi_2^{\prime}\rangle_B$;
with probability $\eta_2$ it is damped to $| \Downarrow \; \rangle_B$;
and finally with probability $\eta_3=1-\eta_1-\eta_2$ it is
transformed in the density matrix $\sigma_B$. In this respect, the map (\ref{rhodibeta2}) 
is similar (but {\em not} equal) to a channel that, with probability $\eta_3$, 
decoheres the input and 
transforms  $|\phi_2\rangle_A$ into $\sigma_B$ while  with probability $1-\eta_3$ applies
to it an amplitude damping channel transformation of quantum efficiency 
$\eta_1/(1-\eta_3)$.

A compact description of~(\ref{rhodibeta2}) 
is obtained by identifying $|\Downarrow \;\rangle_A$, 
$|\Downarrow\;\rangle_B$ with the logical qubit state $|0\rangle$
and $|\phi_2 \rangle_A$, $|\phi_2^\prime\rangle_B$ with $|1\rangle$. 
With this notation, the transformation~(\ref{rhodibeta2}) can be 
expressed as a two-parameter
CPT map,
\begin{eqnarray}
\rho' = {\cal T}_{\eta_1,\eta_2}(\rho) 
\label{NUOVA}
\end{eqnarray}
with Kraus operators given by
\begin{eqnarray}
A_0&=& |0\rangle\langle 0| +\sqrt{\eta_1}\;|1\rangle \langle 1| \;, \nonumber \\
A_1&=& \sqrt{\eta_2}\;|0\rangle \langle 1| \;, \nonumber \\
A_{1+i}&=& \sqrt{\eta_3 \zeta_i} \; |\zeta_{i}\rangle \langle 1| \;.
\label{kraus2}
\end{eqnarray}
Here $|\zeta_{i}\rangle$ are the eigenvectors of $\sigma_B$ associated with
the eigenvalues $\zeta_i>0$: according to Eq.~(\ref{sigmaB}) they are orthogonal
with respect to $|0\rangle$ and $|1\rangle$ and there are at most $k$ of them.

The capacity of the channel ${\cal T}_{\eta_1,\eta_2}$
is derived in Sec.~\ref{s:thecapa1}.

\section{Channel capacities}\label{s:cap}
The quantities we are interested in this paper
are the quantum capacity $Q$, the classical capacity $C$ and
the entanglement assisted capacities $C_E$ and $Q_E$. 
The quantum capacity $Q$  measures the maximum amount of quantum 
information that can be reliably transmitted though the map~$\cal M$ 
per channel use~\cite{SETHQ}.
Intuitively, this quantity is related with the dimension of the largest 
subspace of the multi-uses input Hilbert space which does not decohere 
during the communication process. The value of $Q$ (in qubits per channel uses) 
can be computed as 
\begin{eqnarray} Q \equiv \sup_n \; Q_n/n
\label{quantum}
\end{eqnarray}
with
\begin{eqnarray} 
Q_n \equiv \max_{\rho\in {\cal H}^{\otimes n}} 
\Big\{ S({\cal  M}^{\otimes n} (\rho))-
S(({\cal M}^{\otimes n}
\otimes \openone_{anc}) (\Phi))\Big\}\;\label{q}.
\end{eqnarray}
On one hand, the sup in Eq.~(\ref{quantum}) 
is evaluated over $n$ parallel channel uses \cite{NOTAPARALLEL},
where the channel map is described by the  super-operator
which transforms the input states 
$\rho$ of ${\cal H}^{\otimes n}$ into the
output states  ${\cal M}^{\otimes n}(\rho)$. 
On the other hand, for fixed
values of $n$, the maximization in Eq.~(\ref{q}) 
is performed on all possible
input density matrices  $\rho\in{\cal H}^{\otimes n}$.
The quantity in the bracket is the coherent
information of the channel \cite{SETHQ,SCHUMACHER}, 
$S(\rho)=-\mbox{Tr}
[\rho  \log_2 \rho ]$ is the von Neumann entropy and
$\Phi$ is a 
purification \cite{CHUANG}
of $\rho\in{\cal H}^{\otimes n}$ defined in the extended 
space obtained by 
adding an ancillary space ${\cal H}_{anc}$ to 
${\cal H}^{\otimes n}$.

The classical capacity $C$ 
gives the maximum amount of classical information that 
can be reliably 
transmitted through the channel per channel use: here, 
in the
multi-uses scenario, the goal is 
to identify the largest set of orthogonal input messages 
which remain distinguishable (i.e. orthonormal) during the 
propagation.
In this case, the system is not required to preserve the 
phases 
of superpositions of different messages: $C$ 
characterizes thus the ability of the channel $\cal M$
in preserving occupation numbers but not its decoherence 
effects on the transmitted signals.

As in the case of Eq.~(\ref{quantum}), in calculating
the 
classical capacity it is necessary to perform
a maximization over multiple uses of the channel $\cal M$,
i.e. \cite{HSW}
\begin{eqnarray} C \equiv \sup_n \; C_n/n
\label{classical}
\end{eqnarray}
where $C_n$ is the classical capacity of the channel
which can be achieved if the sender is allowed to encode
the information on 
codewords which are entangled only up to $n$-parallel
channel uses. The value of $C_n$ is 
obtained by maximizing the Holevo information
\cite{HOLEVO} at the output of $n$ parallel channel uses, 
over all possible ensembles $\{\xi_k,\rho_k\}$, i.e.
\begin{eqnarray} 
C_n \equiv \max_{\xi_k,\rho_k\in {\cal H}^{\otimes n}} 
\Big\{ S({\cal  M}^{\otimes n} (\rho))-\sum_k \xi_k
S({\cal M}^{\otimes n} (\rho_k))\Big\}\nonumber\;,\\
\label{c}
\end{eqnarray}
with $\{ \xi_k \}$ 
probabilities and $\rho\equiv\sum_k \xi_k \rho_k$
the average message transmitted.
The optimization~(\ref{classical}) could be avoided if the additivity
conjecture of the Holevo information is true 
\cite{SHOR,shorequiv}: in this case in fact the optimal ensembles  
$\{\xi_k,\rho_k\}$ which achieve the maximum in Eq.~(\ref{c}) are
separable with respect to the $n$ parallel uses and
$C$ coincides with $C_1$.

The last capacities we consider 
are the entanglement-assisted classical
capacity $C_E$ and its quantum counterpart $Q_E$ 
\cite{BENNETT1,BEN}.
These quantities
give, respectively, the maximum amount of
classical or quantum information that 
can be sent reliably through the channel per
channel use, assuming that the sender and the receiver share 
infinite prior entanglement.
To calculate $C_E$ it is not requested to perform a 
regularization
over parallel channel uses as in the case of 
Eqs.~(\ref{quantum}) 
and (\ref{classical}). Here instead one has 
to maximize the quantum mutual information for the
single channel use \cite{BENNETT1,BEN,ADAMI}, i.e.
\begin{eqnarray} 
C_E \equiv \max_{\rho\in {\cal H}} 
\Big\{ S(\rho)+ S({\cal M} (\rho))-S(({\cal M}
\otimes \openone_{anc}) (\Phi))\Big\}\label{ce}
\end{eqnarray}
where now $\Phi$ is a purification of 
the input message $\rho\in \cal H$.
The entanglement assisted quantum capacity can then
be obtained as $Q_E =C_E/2$ 
by means of quantum
teleportation \cite{TELE} and superdense coding 
\cite{SUP}.
The relevance of $C_E$ relies on the fact that this quantity
gives a simple upper bound for the other
capacities. Moreover it is conjectured to provide an
equivalence class for quantum channels \cite{BEN}.

In the next sections we will focus on the capacities of the
CPT maps associated with spin chain models introduced in the 
Secs.~\ref{primavariazione} and 
\ref{secondavariazione}.

\subsection{Capacities of the amplitude damping channel}
\label{s:thecapa}

In this section we analyze in detail the 
qubit amplitude damping channel. In particular
we calculate its capacity 
$Q$ showing that in this case the maximization~(\ref{q}) over
parallel uses is not necessary.
Moreover, we derive the capacities $C_E$ and $C_1$ (classical capacity
achieved with unentangled codewords) which were originally given in Ref. \cite{BEN}
without an explicit derivation.

The map ${\cal D}_\eta$ is completely characterized by the parameter $\eta$ and 
can be seen as an instance of the map ${\cal E}_{\eta}$ associated with the 
lossy Bosonic channel~\cite{LOSSY,LOS,MINIMUM} (see App. \ref{s:app1} for details).
In particular ${\cal D}_\eta$ has the useful property
that by concatenating two amplitude damping channel with
quantum efficiencies $\eta$ and $\eta^\prime$, one obtains
a new amplitude damping channel with efficiency $\eta\eta^\prime$,
i.e.
\begin{eqnarray}
{{\cal D}}_{\eta\prime}\left( {\cal D}_{\eta}(\rho)\right) 
= {\cal D}_{\eta\eta^\prime}(\rho)
\label{lossy3}\;,
\end{eqnarray}
which applies for any input state $\rho$.

\paragraph*{Quantum capacity:--} An important 
simplification in 
calculating the quantity~(\ref{quantum}) 
derives by introducing the following
representation of the CPT map ${\cal D}_\eta$,
\begin{eqnarray}
{\cal D}_\eta (\rho) \equiv 
\mbox{Tr}_C [ V \left( \rho \otimes |0
\rangle_C
\langle 0| \right) V^{\dag}]\;,
\label{lossy}
\end{eqnarray}
obtained by adding to the Hilbert space ${\cal H}_A$ of the
input logical qubit $A$ an auxiliary Hilbert space ${\cal H}_C$,
and introducing the unitary operator $V$ which in  the
computational basis $\{ |0 0 \rangle, 
|0 1 \rangle,
|1 0
\rangle,|1 1 \rangle \}$
of ${\cal H}_A\otimes  {\cal H}_{C}$ is given by the $4\times4$ matrix,
\begin{eqnarray}
V \equiv \left( 
\begin{array}{cccc} 
1 & 0 &0 & 0 \\
0 & \sqrt{\eta} & \sqrt{1-\eta} & 0 \\
0& - \sqrt{1-\eta}&\sqrt{\eta}&0 \\
0&0&0&1\end{array}
\right)\label{unitary}
\;.
\end{eqnarray}
In Eq.~(\ref{lossy}), $\mbox{Tr}_C[\cdots]$ is the partial trace
over the elements of the auxiliary space ${\cal H}_C$. 
The complementary channel $\tilde{\cal D}_\eta$ 
of ${\cal D}_\eta$ is defined by replacing
this operation with the partial trace over ${\cal H}_A$, i.e.
\cite{DEVETAK}
\begin{eqnarray}
\tilde{{\cal D}}_\eta (\rho) 
\equiv \mbox{Tr}_A [ V \left( \rho \otimes |0  
\rangle_C
\langle 0 | \right) V^{\dag}]\;.
\label{lossy1}
\end{eqnarray}
Upon a swapping operation $S$ which transforms 
$A$ into $C$ and vice-versa,
the transformation of Eq.~(\ref{lossy1}) 
can be seen as a mapping from
the Hilbert space ${\cal H}_A$ into itself.
Moreover, by direct calculation one can verify that
\begin{eqnarray}
\tilde{{\cal D}}_\eta (\rho) = 
S \; {{\cal D}}_{1-\eta} (\rho) \; S 
\label{lossy2}\;.
\end{eqnarray}
Using the composition rule of Eq.~(\ref{lossy3})
it is thus possible to show that, for
$\eta\geqslant 0.5$, one has \cite{nota2}
\begin{eqnarray}
\tilde{{\cal D}}_\eta (\rho) =
  S \; {\cal D}_{(1-\eta)/\eta}
\left({{\cal D}}_{\eta} (\rho)\right) \; S 
\label{lossy4}\;.
\end{eqnarray}
[The quantum capacity in the case $\eta<0.5$ is simple to
compute and will be
discussed at the end of this section].
This relation shows that for 
the channel ${\cal D}_{\eta}$ is degradable,
i.e. there exists a
CPT map defined by the super-operator
$ S \; {\cal D}_{(1-\eta)/\eta}(\cdots)\; S $ which connects
the output state 
${{\cal D}}_{\eta} (\rho)$ with the output state
$\tilde{{\cal D}}_\eta (\rho)$.
According to a theorem proved
by Devetak and Shor \cite{DEVETAK}, 
this condition guaranties that
the sup in Eq.~(\ref{quantum}) is 
achieved for $n=1$ (single channel use): in
other words, Eq.~(\ref{lossy4}) guarantees the
additivity of the coherent information
 for the channel ${\cal D}_\eta$.
For $\eta\geqslant 0.5$ 
the quantum capacity of $\cal D_{\eta}$ 
derives thus by solving the 
maximization~(\ref{q}) for $n=1$.

\begin{figure}[t]
\begin{center}
\epsfxsize=.99\hsize\leavevmode\epsffile{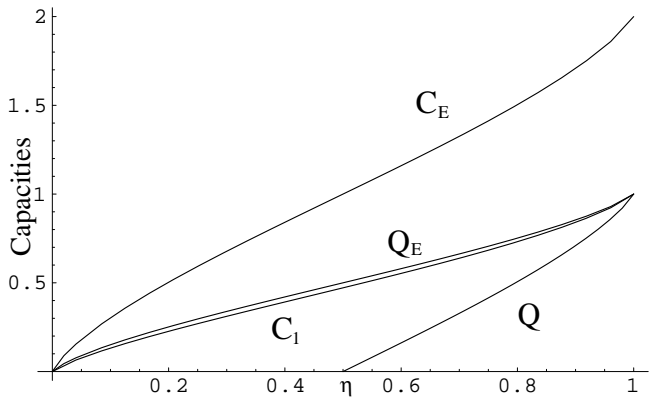}
\end{center}
\caption{Plot of the capacities (in bits per channel uses) 
of the qubit amplitude
damping channel ${\cal D}_\eta$ as a function of the
noise parameter $\eta$. The curves $Q$, $C_E$ and $C_1$
represent respectively, the quantum capacity, entanglement
assisted classical capacity, and the classical capacity
achievable with unentangled encodings: they 
have been obtained
 by solving numerically the maximizations 
of Eqs.~(\ref{qcalc}), (\ref{finito2}) and (\ref{up}).
For $\eta=1$ (no noise limit) both $Q$ and $C_1$
give one qubit per bit for channel use, while
as a consequence of the superdense coding effect 
\cite{SUP} $C_E$ gives two bits per channel uses.  
For $\eta=0.5$ we have, respectively, $Q=0$, $C_E=1$, and
$C_1=0.4717$ (this result is in agreement
with what found in Ref. \cite{SCHUMI}).
As a consequence of the
non cloning theorem \cite{SMOLIN,LOSSY} 
for $\eta<0.5$ the
quantum capacity $Q$ nullifies. 
The curve $Q_E$ represents the 
entanglement assisted quantum capacity and is obtained
by simply dividing by two the values of $C_E$. The
classical capacity $C$ of the channel is lower bounded by 
$C_1$ and upper bounded
by $1$ (maximum entropy of a bit) and by $C_E$. If the
additivity conjecture \cite{shorequiv} applies to this
channel then $C=C_1$.}
\label{f:fig1}
\end{figure}

\begin{figure}[t]
\begin{center}
\epsfxsize=.99\hsize\leavevmode\epsffile{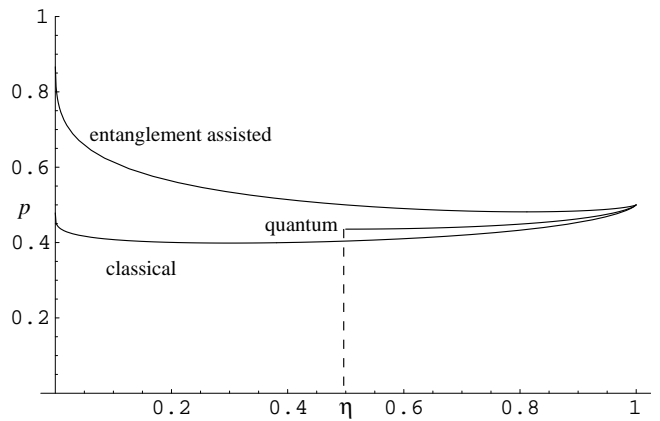}
\end{center}
\caption{Plot of the optimal populations $p$ associated with
the state 
$| 1 \rangle$
which provide the capacities
of the amplitude damping channel  ${\cal D}_\eta$, 
as a function of the transmissivity $\eta$.
The upper curve refers to the entanglement
assisted capacities $C_E$ and $Q_E$: for given $\eta$
it has been obtained by finding that value of
$p$ which maximizes the function at the 
right-hand-side of Eq. (\ref{finito2}).
The lower curve refers to the classical capacity $C_1$
through Eq.~(\ref{up}). Finally the intermediate curve refers
to the quantum capacity $Q$ through Eq.~(\ref{qcalc}):
this function is not defined for $\eta<0.5$
because for this value the $Q$ is null (see 
Fig.~\ref{f:fig1}). For $\eta\geqslant 0.5$ the optimal 
population $p$ of the quantum capacity is bigger than 
the corresponding  population
of $C_1$ and lower than that of $C_E$.}
\label{f:fig2}
\end{figure}

In the computational basis 
$\{ |0\rangle , | 1\rangle
\}$ of ${\cal H}_A$, the most general input state of the 
map ${\cal D}_{\eta}$ can
be parametrized as follows
\begin{eqnarray}
\rho \equiv \left( 
\begin{array}{cc} 
1-p & \gamma^*\ \\
\gamma & p 
\end{array}
\right)\;,
\label{rhoin}
\end{eqnarray}
where $p\in[0,1]$ is the population
associated with the state $|1 \rangle$
and $|\gamma|\leqslant \sqrt{(1-p)p}$ is a coherence
term.
A purification $\Phi\equiv|\Phi\rangle\langle \Phi|$
of $\rho$ is then obtained by introducing an
ancillary qubit system ${\cal H}_{anc}$ 
and considering the state
\begin{eqnarray}
|\Phi\rangle \equiv \sqrt{1-p}\; 
|0 \rangle \otimes | R_{0}\rangle
+ \sqrt{p}\; |1 \rangle \otimes | R_{1}\rangle
\label{puri}\;,
\end{eqnarray}
with $| R_{0,1}\rangle$ unit vectors of 
${\cal H}_{anc}$ such
that 
\begin{eqnarray}
\langle R_{0}| R_{1}\rangle = \gamma/
\sqrt{(1-p)p}
\label{puri1}\;.
\end{eqnarray}
From the Kraus decomposition (\ref{kraus}) one can
verify that the map ${\cal D}_{\eta}$ 
transforms this  state into the output
\begin{eqnarray}
{\cal D}_{\eta} (\rho) = \left( 
\begin{array}{cc} 
1-\eta\, p  & \sqrt{\eta} \, \gamma^* \\
\sqrt{\eta}\,\gamma & \eta\, p 
\end{array}
\right)\;.
\label{rhoout}
\end{eqnarray}
The matrix~(\ref{rhoout}) has eigenvalues 
\begin{eqnarray}
\lambda_{\pm}(\eta) \equiv 
\left(1 \pm \sqrt{(1- 2\,\eta\, p)^2
+4\,\eta\, |\gamma|^2} \right)/2
\label{eigen}\;,
\end{eqnarray}
which gives an output von Neumann entropy equal to
\begin{eqnarray}
S({\cal D}_{\eta} (\rho)) =
H_2 (\lambda_+(\eta))\;,
\label{outentropy}
\end{eqnarray}  
with $H_2$ the binary entropy function 
defined in Eq.~(\ref{bin}).
Analogously, by applying the map
$({\cal D}_{\eta} \otimes\openone_{anc})$ to
$\Phi$ we get the state
described by the $4\times 4$ complex matrix
of Eq.~(\ref{rhoout1}), which has entropy
equal to
\begin{eqnarray}
S(({\cal D}_{\eta} \otimes\openone_{anc})  
(\Phi))  &=& H_2 (\lambda_+(1-\eta))\;.
\label{exentropy1}
\end{eqnarray}
The quantity we need to maximize to obtain $Q$
is hence given by 
\begin{eqnarray}
J(p,|\gamma|^2) \equiv H_2 (\lambda_+(\eta)) -  H_2 (\lambda_+(1-\eta))
\label{COHERENTINFO}
\end{eqnarray}
which 
depends from $\rho$ only through
the parameters $p\in[0,1]$ 
and $|\gamma|^2\in[0,(1-p)p]$. 
As discussed in App.~\ref{s:app2} the 
maximization~(\ref{quantum}) is achieved by
choosing $\gamma=0$, which gives a quantum capacity
equal to
\begin{eqnarray}
Q&\equiv&
\max_{p\in[0,1]} \; \Big\{ \;
H_2 (\eta\, p) - H_2((1-\eta)\, p)\label{qcalc}\; \Big\}\;.
\end{eqnarray}
For any given $\eta\geqslant0.5$ this expression
has been solved numerically and the results are plotted in
Fig.~\ref{f:fig1}. In Fig.~\ref{f:fig2}, instead, we 
have reported, as function of the parameter $\eta$,
 the optimal value of the population
$p$ which provides the maximum
of the right-hand-side term of Eq.~(\ref{qcalc}).

Let us now consider the low transmissivity regime \mbox{$
\eta<0.5$}.
In this case, the same non-cloning argument given in
Ref.~\cite{SMOLIN} for the erasure channel 
and in Ref.~\cite{LOSSY} for the
lossy Bosonic channel, can be used to
prove that the quantum capacity $Q$ of ${\cal D}_\eta$ 
nullifies.
An alternative proof of this fact, can be obtained
by noticing that the composition rule~(\ref{lossy3})
implies that the quantum capacity of the channel
${\cal D}_\eta$ is
an increasing function of the transmissivity $\eta$.
The thesis then follows from the fact that for $\eta=0.5$
the right-hand-side term of Eq.~(\ref{qcalc}) nullifies.

\paragraph*{Entanglement assisted capacity:--} 
To calculate this capacity we need to perform the
maximization of Eq.~(\ref{ce}) for all possible
input states $\rho$.
The quantum mutual information of the 
channel can be obtained by summing the coherent information $J(p,|\gamma|^2)$
of Eq.~(\ref{COHERENTINFO}) to the
input entropy of the message, i.e. using
the parametrization introduced in the previous section,
\begin{eqnarray}
&&I(p,|\gamma|^2) \equiv J(p,|\gamma|^2)\label{mutu}
\\ \nonumber  
&&\qquad\qquad + H_2 \left(\frac{1 + \sqrt{(1- 2 p)^2
+4 \, |\gamma|^2}}{2} \right)\;.
\end{eqnarray}
According to the property
{\bf 1)} of App. \ref{s:app21} 
the last term in the right-hand-side of
this expression is a decreasing function of $|\gamma|^2$, 
i.e. it is maximum for $\gamma=0$. 
From the previous section, we know that the same
property applies also to $J(p,|\gamma|^2)$:
we can thus 
conclude that, for any $p\in[0,1]$, the function
$I(p,|\gamma|^2)$ 
achieves its maximum value for $\gamma=0$. 
In other words, we can compute the entanglement assisted
capacity $C_E$ as
\begin{eqnarray} 
C_E&\equiv&
\max_{p\in[0,1]} \; \Big\{ \; H_2( p) +
H_2 (\eta\, p) - H_2((1-\eta)\, p)\; \Big\}\;.
\nonumber\\
\label{finito2}
\end{eqnarray}
This maximization can now be solved numerically:
the resulting plot is given in Fig.~\ref{f:fig1}.
The optimal $p$'s that saturate the maximization
of Eq.~(\ref{finito2}) are reported in Fig.~\ref{f:fig2}.

\paragraph*{Classical capacity with unentangled encodings:--}

In the case of 
the lossy Bosonic channel ${\cal E}_\eta$
with
constrained input average photon number
the additivity property of the Holevo information has been proved \cite{LOS}.
Unfortunately, this  derivation 
relies on some specific properties of the 
coherent input state of the Bosonic channel:
even though ${\cal E}_\eta$ and ${\cal D}_\eta$ are strongly related 
it is hence difficult to use the
result of \cite{LOS} to establish the additivity conjecture
for the qubit amplitude damping channel (see also App. \ref{s:app1}).
Here we will not discuss further this problem 
and simply we will 
focus on the capacity $C_1$, which measures the
maximum amount of classical information
that can be reliably transmitted using only
encodings that are not entangled over parallel
channel uses \cite{HSW,SHOR,shorequiv}. 
The quantity $C_1$ is a lower
bound for $C$ and coincides with it
provided the additivity conjecture holds.
The classical capacity $C_1$ can be calculated by solving the
maximization of Eq.~(\ref{c}) in the case of $n=1$.
Consider the ensemble of messages where with
probability $\xi_k$ the channel is prepared in the
input state
\begin{eqnarray}
\rho_k \equiv \left( 
\begin{array}{cc} 
1-p_k & \gamma_k^*\ \\
\gamma_k & p_k 
\end{array}
\right)\;,
\label{rhoink}
\end{eqnarray}
with $p_k$ and $\gamma_k$ defined as in Eq.~(\ref{rhoin}).
Using the result of the previous section we can express
the associated Holevo information as
\begin{eqnarray}
&&\chi 
\equiv H_2 \left(\frac{1 + \sqrt{(1- 2 \,\eta\,p)^2
+4 \,\eta\, |\gamma|^2}}{2} \right)\label{holevoinf}
\\
&&\qquad -\sum_k \xi_k
H_2 \left(\frac{1 + \sqrt{(1- 2 \,\eta\,p_k)^2
+4 \,\eta\, |\gamma_k|^2}}{2} \right)\;,
\nonumber
\end{eqnarray}
where now $p= \sum_k \xi_k p_k$ and 
$\gamma\equiv \sum_k \xi_k \gamma_k$  are the 
parameters~(\ref{rhoin}) associated  with the
average input message $\rho \equiv \sum_k \xi_k \rho_k$.
According to Eq.~(\ref{c}) 
the capacity $C_1$ is obtained by maximizing $\chi$
over all possible choices of $p_k$, $\gamma_k$
and $\xi_k$. To solve this problem we first
derive an upper bound for $C_1$ and then 
we show that there exist an encoding $p_k$, $\gamma_k$
and $\xi_k$ which
achieves such an upper bound.

From the property {\bf 1)} of the binary 
entropy (\ref{bin}) given in App. \ref{s:app21}, we 
can maximize the first term in the right-hand-side 
of Eq.~(\ref{holevoinf}) by choosing $\gamma=0$.
Moreover, one has
\begin{eqnarray}
&&\sum_k \xi_k 
H_2 \left(\frac{1 + \sqrt{(1- 2 \,\eta\,p_k)^2
+4 \,\eta\, |\gamma_k|^2}}{2} \right) \label{dis}\\
&&\qquad \geqslant 
\sum_k \xi_k 
H_2 \left(\frac{1 + \sqrt{1- 4 \,\eta\,(1-\eta) p_k^2}}
{2} \right) \nonumber \\
&&\qquad \geqslant 
H_2 \left(\frac{1 + 
\sqrt{1- 4 \,\eta\,(1-\eta) (\sum_k \xi_k
p_k) ^2}}
{2} \right)\;.
\nonumber
\end{eqnarray}
where the first inequality derives from the property
{\bf 1)} of $H_2(z)$ and from the fact that 
$|\gamma_k|^2\leqslant (1-p_k)p_k$, while the
second inequality is consequence of the property
{\bf 2)}.
Replacing the above relation
in Eq.~(\ref{holevoinf}) we obtain an
upper bound for $\chi$ which does not
depend on $\gamma_k$ and which
depends on $\xi_k$ and $p_k$  only
through  $p=\sum_k \xi_k p_k$.
By maximizing this expression over all possible
choices of the variable $p$ we get the
following upper bound of $C_1$,
\begin{eqnarray}
C_1 &\leqslant& \max_{p\in[0,1]}
\Big\{ H_2 \left(\eta \, p \right)\label{up}\\
&&\qquad -H_2 \left(\frac{1 + 
\sqrt{1- 4 \,\eta\,(1-\eta) \,p ^2}}{2}
\right) \Big\}
\;.\nonumber
\end{eqnarray}
The right-hand-side term of this
inequality is indeed the value of $C_1$. This 
can be show by noticing that
for any $p\in[0,1]$ and $d>1$, the parameters 
\begin{eqnarray}
\xi_k=1/d \qquad p_k=p \nonumber \\
\gamma_k=e^{2\pi i k/d} \; \sqrt{(1-p)p}\;,
\label{encoding}
\end{eqnarray}
with $k=1,\cdots,d$, produce a Holevo information
$\chi$ of Eq.~(\ref{holevoinf}) which is coincident
with the quantity in the brackets on the right-hand-side
of Eq.~(\ref{up}).
The quantities~(\ref{encoding}) provide hence optimal
encoding strategies for $C_1$. On one hand,
any ensemble element $\rho_k$ of this encoding
has maximum absolute value of the
coherence term $\gamma_k$: this minimizes the negative
term of the Holevo information. On the other hand,
the average message $\rho=\sum_k \xi_k \rho_k$ has
minimum value $|\gamma|$, i.e.
\begin{eqnarray}
\gamma=\sum_{k=1}^d \xi_k \gamma_k =
\sqrt{(1-p)p}\sum_{k=1}^d  e^{2 \pi i k/d}/d = 0\;,
\label{ggamma} 
\end{eqnarray}
which maximizes the positive contribution to the
Holevo information.
This property of the channel ${\cal D}_\eta$ is a
common feature of many other channels whose 
classical capacity $C$ has been solved \cite{LOS,KING}.
The value of $C_1$ obtained by maximizing the 
right-hand-side term of Eq.~(\ref{up}) has been plotted in 
Fig.~\ref{f:fig1}, while the optimal $p$'s are plotted
in Fig.~\ref{f:fig2}.

\subsection{Capacities of the channel ${\cal T}_{\eta_1,\eta_2}$}
\label{s:thecapa1}

In this section we analyze in details the 
CPT map  ${\cal T}_{\eta_1,\eta_2}$ of Eq.~(\ref{NUOVA})
associated with the spin chain communication line of Sec.~\ref{secondavariazione}.
In this case we calculate the capacities 
$Q$, $C_E$ and  we provide a lower bound for $C$.

The map  ${\cal T}_{\eta_1,\eta_2}$ is described by the 
positive parameters $\eta_1$ and $\eta_2$ of Eqs.~(\ref{etanew1}) and
(\ref{etanew2}) that satisfy the relation $\eta_1+\eta_2\leqslant 1$.
In particular, for $\eta_2=1-\eta_1$, ${\cal T}_{\eta_1,\eta_2}$ reduces to 
an amplitude damping map of transmissivity $\eta_1$,
i.e.  
\begin{eqnarray}
{\cal T}_{\eta_1,1-\eta_1}(\rho)
= {\cal D}_{\eta_1}(\rho)
\label{lossy3lossy}\;,
\end{eqnarray}
for any input state $\rho$.  Moreover the following composition rule applies, 
\begin{eqnarray}
{{\cal T}}_{\eta_1\prime,\eta_2^\prime}\left( {\cal T}_{\eta_1,\eta_2}(\rho)\right) 
= {\cal T}_{\eta_1^{\prime\prime},\eta_2^{\prime\prime}}(\rho)
\label{lossy3lossy3}\;,
\end{eqnarray} 
with $\eta_1^{\prime\prime} \equiv \eta_1 \eta_1^{\prime}$ and
$\eta_2^{\prime\prime}\equiv \eta_2 + \eta_1 \eta_2^{\prime}$.
An interesting way to express ${\cal T}_{\eta_1,\eta_2}$
is finally provided by concatenating two amplitude damping channels with a 
CPT map $\cal P$ that transforms one of the eigenvectors
of the matrix $\sigma_B$ (say $|\zeta_1\rangle$) into $\sigma_B$ itself, i.e.
\begin{eqnarray}
{\cal T}_{\eta_1,\eta_2}(\rho) = {\cal P} ({\cal D}_{{\eta_1}/({1-\eta_2})}^{\prime}
(  {\cal D}_{1-\eta_2}(\rho) ))\;,
\label{prox}
\end{eqnarray}
where the first amplitude damping channel ${\cal D}_{1-\eta_2}$
acts as usual on $|0\rangle$ and
$|1\rangle$, while second one ${\cal D}_{{\eta_1}/(1-\eta_2)}^\prime$
is instead defined on the subspace generated by $|\zeta_1\rangle$ and $|1\rangle$.
The main consequence of Eq.~(\ref{prox}) is that any 
capacity of ${\cal T}_{\eta_1,\eta_2}$ 
cannot be greater than the corresponding capacity of ${\cal D}_{1-\eta_2}$.
In fact, by applying the CPT transformations 
${\cal P}$ and ${\cal D}_{{\eta_1}/({1-\eta_2})}^\prime$ to the output of  
an amplitude damping channel
of transmissivity $1-\eta_2$ one can simulate the corresponding
output of ${\cal T}_{\eta_1,\eta_2}$.

\begin{figure}[t]
\begin{center}
\epsfxsize=.8\hsize\leavevmode\epsffile{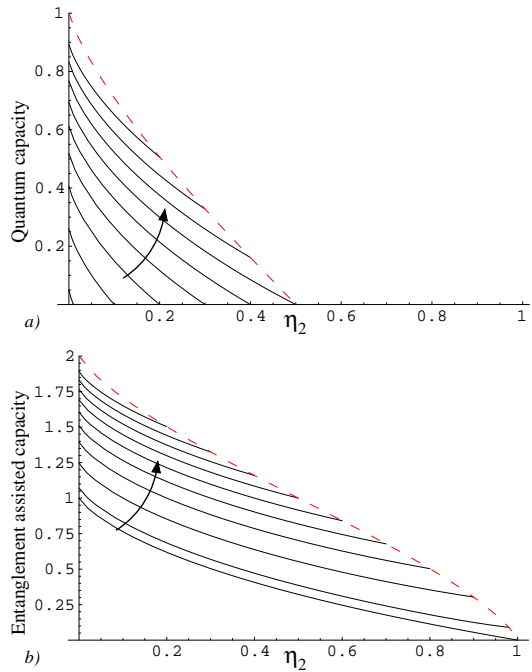}
\end{center}
\caption{Plots of the capacities (in bits per channel uses) of 
the channel ${\cal T}_{\eta_1,\eta_2}$ of Eq.~(\ref{NUOVA})
as a function of $\eta_2\in[0,1-\eta_1]$ obtained for different
$\eta_1$ (this parameter increases from $0$ to $1$ moving along the arrows).
In both graphics the dashed curve is the capacity of an amplitude damping
channel with efficiency $1-\eta_2$ which according to Eq.~(\ref{prox}) provides
an upper bound for the corresponding capacity of  ${\cal T}_{\eta_1,\eta_2}$.
Part {\em a)}: quantum capacity $Q$ obtained by solving numerically the maximization
over $p$ of Eq.~(\ref{qcalc2}). Notice that for $\eta_2>\eta_1$ $Q$ nullifies 
\cite{ARGO}. 
Part {\em b)}, entanglement assisted classical
capacity $C_E$ of Eq.~(\ref{finito22}).}
\label{figura3}
\end{figure}

\paragraph*{Quantum Capacity:--}
As in the case of ${\cal D}_\eta$ we can prove that 
${\cal T}_{\eta_1,\eta_2}$ is degradable when its quantum
capacity is not null.
In fact,  as in Eq.~(\ref{lossy}) 
define the unitary operator $V$ of the extended
Hilbert space ${\cal H}_A\otimes {\cal H}_C$ such that
for any $\rho$ of ${\cal H}_A$ 
\begin{eqnarray}
{\cal T}_{\eta_1,\eta_2} (\rho) \equiv 
\mbox{Tr}_C [ V \left( \rho \otimes |0
\rangle_C
\langle 0| \right) V^{\dag}]\;.
\label{lossynew2}
\end{eqnarray}
An example of $V$ can be obtained by introducing the
following vector of ${\cal H}_A\otimes {\cal H}_C$
\begin{eqnarray}
|\Phi_\sigma\rangle \equiv \sum_i \sqrt{\zeta_i} |\zeta_i\rangle_A 
\otimes |\zeta_i\rangle_C
\label{purifNEW} \;,
\end{eqnarray}
where $|\zeta_i\rangle_A\in{\cal H}_A$ are the eigenvectors of 
$\sigma_B$ introduced in Eq.~(\ref{kraus2})
while   $|\zeta_i\rangle_C$ is an orthonormal set of states of ${\cal H}_C$ 
that are orthogonal
to $|0\rangle_C$ and $|1\rangle_C$.
The state $|\Phi_\sigma\rangle$ is a purification of $\sigma_B$ on  
${\cal H}_A\otimes {\cal H}_C$. The unitary operator $V$ can 
now be chosen to be the identity everywhere 
but on the subspace of ${\cal H}_A\otimes {\cal H}_C$ generated by
the orthonormal vectors  $\{ |00\rangle,|01\rangle,|10\rangle,|\Phi_\sigma\rangle \}$.
On this subset we define $V$ to have the matrix representation  
\begin{eqnarray}
V \equiv \left( 
\begin{array}{ccccc} 
 \frac{1+\sqrt{\eta_1}-\eta_2}{1+\sqrt{\eta_1}} & \sqrt{\eta_2} 
&-  \frac{\sqrt{\eta_2 \eta_3}}{1+\sqrt{\eta_1}} \\
 -\sqrt{\eta_2}& \sqrt{\eta_1} &-\sqrt{\eta_3} \\
-  \frac{\sqrt{\eta_2 \eta_3}}{1+\sqrt{\eta_1}} & \sqrt{\eta_3}& 
\frac{\eta_2 + \eta_3 \sqrt{\eta_1}}{1-\eta_1}
\end{array}
\right)\label{unitary2}
\;,
\end{eqnarray}
with $\eta_3=1-\eta_1-\eta_2$.
The complementary map $\tilde{\cal T}_{\eta_1,\eta_2}$ 
is finally obtained by substituting
in~(\ref{lossynew2}) the trace over $C$ with the trace on $A$ \cite{DEVETAK}.
From Eqs.~(\ref{lossy3lossy}) and (\ref{lossy3lossy3})
one can easily verify that for $\eta_1\geqslant \eta_2$ the following relation applies 
for any input $\rho$,
\begin{eqnarray}
\tilde{\cal T}_{\eta_1,\eta_2} (\rho) 
=  S\; {\cal D}_{\eta_2/\eta_1} ({\cal T}_{\eta_1,\eta_2} (\rho))
\;S
\label{degrado2}
\end{eqnarray}
with $S$ the swapping operation which transforms $A$ in $C$.
Since ${\cal D}_{\eta_2/\eta_1}$ is CPT the above 
equation shows that for $\eta_1\geqslant \eta_2$ the map  ${\cal T}_{\eta_1,\eta_2}$ is
degradable: the quantum capacity of this channel can be hence computed from Eq.~(\ref{q})
for $n=1$. 
A straightforward generalization of the qubit amplitude damping channel analysis
shows that Eq.~(\ref{COHERENTINFO}) still applies by replacing $\lambda_{\pm}(\eta)$
of Eq.~(\ref{eigen}) with
\begin{eqnarray}
&&\lambda_{\pm}(\eta_1,\eta_2) \equiv \frac{1-\eta_3 p}{2} \label{eigennew2}\\
&& \quad \quad \times \left[1 \pm \sqrt{\left(1- \frac{2\,\eta_1\, p}{1-\eta_3p}\right)^2
+\frac{4\,\eta_1\, |\gamma|^2}{(1-\eta_3 p)^2}} \; \right]
\nonumber \;.
\end{eqnarray}
The quantum capacity of ${\cal T}_{\eta_1,\eta_2}$ becomes hence
\begin{eqnarray}
&&Q\equiv
\max_{p\in[0,1]} \; \Big\{ \;
(1-\eta_3 p)\label{qcalc2} \\
&&\times \left[ \; H_2 \left(\frac{1-(1-\eta_2) p}{1-\eta_3 p}\right) - 
H_2\left(\frac{1-(1-\eta_1) p}{1-\eta_3 p}\right) \; \right] \nonumber \; \Big\}\;.
\end{eqnarray}
Notice that as in the case of Eq.~(\ref{qcalc}) the maximization over the input
parameter $\gamma$ of Eq.~(\ref{rhoin}) has been saturated by setting $\gamma=0$
(the proof goes as in  Eq.~(\ref{qcalc})). A plot of $Q$ as a function of
$\eta_1$ and $\eta_2$ is reported in Fig.~\ref{figura3} by solving numerically the
maximization on $p$. 
The above results do not apply for $\eta_1 \leqslant \eta_2$:
in this case in fact ${\cal D}_{\eta_2/\eta_1}$ is not CPT. However, 
a non-cloning argument can be
used to prove that in this parameters region, 
the quantum capacity of  ${\cal T}_{\eta_1,\eta_2}$ nullifies \cite{ARGO}.

\paragraph*{Entanglement assisted capacity:--}
The analysis of $C_E$ proceeds as in the case of the
amplitude damping channel. Here Eq.~(\ref{finito2}) is
replaced by
\begin{eqnarray} 
&&C_E\equiv 
\max_{p\in[0,1]} \; \Big\{ \;  H_2( p) + 
(1-\eta_3 p)\label{finito22} \\
&&\times \left[ \; H_2 \left(\frac{1-(1-\eta_2) p}{1-\eta_3 p}\right) - 
H_2\left(\frac{1-(1-\eta_1) p}{1-\eta_3 p}\right) \; \right] \nonumber \; \Big\}\;.
\end{eqnarray}
A numerical plot of this expression is given in Fig.~\ref{figura3}.

\paragraph*{Lower bound for $C_1$:--}
The analysis of $C_1$ for the channel ${\cal T}_{\eta_1,\eta_2}$ is slightly more difficult 
than that for ${\cal D}_\eta$ as the convexity properties used in Eq.~(\ref{dis})
do not hold in this case.
Here thus we gives a lower bound for $C_1$ obtained by 
assuming the encoding Eq.~(\ref{encoding}),
\begin{eqnarray}
C_1 &\geqslant& \max_{p\in[0,1]} 
\Big\{ \; 
(1-\eta_3 p)\; \Big[ \; H_2 \left(\frac{1-(1-\eta_2) p}{1-\eta_3 p}\right)
\nonumber \\
&&- H_2\left(\frac{1}{2}+\frac{1}{2}\sqrt{1-\frac{4\eta_1\eta_2 p^2}{(1-\eta_3 p)^2}
}\;\right) \; \Big] \; \Big\}\;.
\label{up2}
\end{eqnarray}

\section{Conclusions}\label{s:tre}
In the previous section we calculated the
capacities $Q$, $C_E$ and $C_1$ for the qubit 
amplitude damping channel ${\cal D}_\eta$ and for the channel
${\cal T}_{\eta_1,\eta_2}$. As discussed in
Sec. \ref{s:solvable}, by identifying the parameters 
$\eta$, $\eta_1$ and $\eta_2$ with the quantities of 
Eqs.~(\ref{etanew}), (\ref{etanew1}) and
(\ref{etanew2}) 
we can use these results to analyze the correlations between
distant points of the chain.
A detailed analysis of the fidelity for state transfer~(\ref{etanew}) 
in the case of $k=1$ and uniform coupling $J_{i,j}$ is given in Ref.~\cite{BOSE}.
Such a paper solves the dependence of $\eta$ from the evolution time and from the 
length of the chain. In particular in \cite{BOSE} 
it is shown that the  maximum value of $\eta$
drops with the distance between the encoding spin $A$ (at the site $r$) 
and the decoding spin $B$ (at site $s$)  as  $|r-s|^{-2/3}$.
Since according to Sec.\ref{s:thecapa} 
the quantum capacity $Q$ vanishes
when  the transmissivity is $\leqslant 0.5$,
this implies that long chains are not suitable 
to directly transmit quantum information using the scheme of Sec.~\ref{primavariazione}
with $k=1$.
Even though the communication scenarios described by means of Eq.~(\ref{mappa}) are
incomplete (see the discussion of Sec.~\ref{s:uno}) it is 
interesting to see in which way the above difficulties can be overcome.
One possibility is to optimize the values of $\eta$ 
by tailoring the interaction $J_{i,j}$ 
between the spins as proposed in Ref.~\cite{DATTA}. Alternatively
one can use registers with $k>1$ and then 
optimizing the value
of $\eta$ of Eq.~(\ref{etanew}) by means an appropriate choice \cite{OSBORNE} of the
coefficients $c_j$ of Eq.~(\ref{PHI1}).

Yet a different approach would require the partition of the chain in smaller 
segments so to transfer information faithfully along the chain through swaps 
between neighboring segments analogously to the ideas of quantum repeaters introduced in 
\cite{BRIGEL}. Breaking the chain can be 
achieved by applying locally external magnetic fields. A combination of 
a time dependent control of part of the chain together with 
perfect transmission of small segments may lead to an improvement of the 
performances of spin chains to transport quantum information.  
On the other hand, the quantum capacity of the chain can be
boosted by giving access the sender and receiver to
a free two-way classical communication line \cite{BEN,DIST}. In this case, 
the ability of the channel in transmitting quantum signal can be 
increased by means of entanglement distillation protocols and
teleportation. For instance (as noticed also in \cite{BOSE}) since any two qubits 
entanglement state is distillable \cite{HORO},
the two-way quantum capacity $Q_{2\mbox{\tiny{-way}}}$ 
of the chain is strictly greater
than zero also when $Q$ is null. To compute the exact value
of  $Q_{2\mbox{\tiny{-way}}}$ one needs to find the optimal 
distillation protocol for ${\cal M}$: unfortunately
this is quite a challenging task.

\acknowledgments
 
This work was supported by the European Community under contract IST-SQUIBIT,
IST-SQUBIT2, and RTN-Nanoscale Dynamics.

\appendix

\section{Derivation of Eq.~(\ref{rhodibeta2})}\label{a:postpo}
The Hamiltonian~(\ref{hamilto}) preserves the total spin component along
the $z$ axis. The set of two-spin up states 
$|{ j}, { \ell} \rangle$ is hence transformed into
itself with unitary matrix elements
\begin{eqnarray}
f_{j\ell,j^\prime\ell^\prime}(t)\equiv
\langle { j},{ \ell} | e^{-i 
H t/\hbar}|
{ j^\prime},{ \ell^\prime}\rangle \label{effeNEW}\;.
\end{eqnarray}
Using this transformation it is possible to show that 
at time $t$ the spin chain state becomes
\begin{eqnarray}
|\Psi(t)\rangle \equiv \alpha |\Downarrow \; \rangle + \beta
\sum_{j>\ell=1}^N d_{j,\ell}(t) |{ j },{  \ell }
\rangle \;,
\label{OUTPUT2}
\end{eqnarray}
where, for $j>\ell=1,\cdots,N$, 
\begin{eqnarray}
d_{j,\ell}(t) \equiv \sum_{j^\prime>\ell^\prime=1}^k 
d_{j^\prime\ell^\prime} f_{j\ell,j^\prime\ell^\prime}(t)
\label{dout}
\end{eqnarray}
are the time evolved of the coefficient $d_{j,\ell}$ of Eq.~(\ref{PHI2}).
Equation~(\ref{OUTPUT2})  shows that only the fraction 
\begin{eqnarray}
\eta_1 \equiv \sum_{j>\ell=N-k+1}^N 
\left| d_{j,\ell}(t) \right |^2
\label{etanew1}\end{eqnarray}
of $|\Psi(t)\rangle$ has two spins up in $B$. Analogously
\begin{eqnarray}
\eta_2 &\equiv& \sum_{j>\ell= 1}^{N-k} 
\left|d_{j,\ell}(t)\right|^2 \;,
\label{etanew2}
\end{eqnarray}
is the probability of $|\Psi(t)\rangle$ having both two spins up 
outside of $B$, while
\begin{eqnarray}
\eta_3\equiv 1-\eta_1-\eta_2 &=& \sum_{\ell= 1}^{N-k}\sum_{j=N-k+1}^{N} 
\left|d_{j,\ell}(t)\right|^2\;,
\label{etanew3}
\end{eqnarray}
is the probability of having one spin up in $B$ and the other outside of $B$.
By taking the partial trace of~(\ref{OUTPUT2}) 
over the first $N-k$ spins of the chain we obtain the
state of the quantum
memory $B$ of Eq.~(\ref{rhodibeta2}). In such an expression,
\begin{eqnarray}
|\phi_2^{\prime}\rangle_B \equiv \sum_{j>\ell =N-k+1}^{N} 
d_{j,\ell}(t)|j, {\ell} \rangle /\sqrt{\eta_1} \;.
\label{PHIPRIME2}
\end{eqnarray}
is a rotation of $|\phi_2\rangle_A$ that can be compensated for at the
decoding stage. On the other hand, $\sigma_B$ of Eq.~(\ref{rhodibeta2}) 
is a density matrix of 
$B$ formed by states with one-spin-up vectors, i.e.
\begin{eqnarray}
\sigma_B \equiv \sum_{\ell=1}^{N-k}
\eta_3(\ell) \; |\phi_1(\ell)\rangle \langle \phi_1(\ell)|/\eta_3\;,
\label{sigmaB}
\end{eqnarray}
with 
\begin{eqnarray}
|\phi_1(\ell)\rangle_B &\equiv& \sum_{j  =N-k+1}^{N} 
d_{j,\ell}(t)|j, {\ell} \rangle / \sqrt{\eta_3(\ell)} \nonumber \\
\eta_3(\ell)&\equiv&\sum_{j =N-k+1}^N |d_{j,\ell}(t)|^2 \;.
\label{PHIPRIME1elle}
\end{eqnarray}
Notice that the rank of $\sigma_B$ is at most equal to  
the maximum number  of orthogonal one-spin up states of $B$, that is $k$. Moreover 
the support of $\sigma_B$ is clearly orthogonal to $|\Downarrow \,\rangle_B$ and
$| \phi_2^\prime\rangle_B$.

\section{Relation with the 
Bosonic lossy channel}\label{s:app1}
In the lossy Bosonic channel ${\cal E}_\eta$ \cite{LOSSY,LOS,MINIMUM}
an input 
Bosonic mode
described  by the annihilation
operator $a$ interacts, through a beam splitter 
of transmissivity $\eta$, 
with the vacuum state $|\O \rangle_b$ of an
 external Bosonic mode described by the annihilation 
operator $b$. Any input state
$\rho$ of the mode $a$ is hence transformed by this map 
according to the equation 
\begin{eqnarray}
{\cal E}_\eta (\rho) \equiv 
\mbox{Tr}_b [ U \left( \rho \otimes |\O
\rangle_b
\langle \O| \right) U^{\dag}]
\label{lossyAPPENDIX}
\end{eqnarray}
where the trace is performed over the 
external mode $b$ and where
$U$ is the beam splitter unitary operator defined by
\begin{eqnarray}
U^{\dag} a\; U &=& \sqrt{\eta}\;a + \sqrt{1-\eta} \;b \\
U^{\dag} b \; U &=& \sqrt{\eta}\;b -\sqrt{1-\eta} \;a
\label{beamAPPENDIX}\;.
\end{eqnarray}
By restricting the inputs 
$\rho$ to the Hilbert space spanned by the vacuum state and
the one photon
Fock state the map~(\ref{lossy}) has
the same Kraus decomposition~(\ref{kraus}) 
of the qubit map ${\cal D}_\eta$.
Some capacities of the channel $\cal{E}_\eta$ have been solved
under constrained average input photon number: 
the classical capacity $C$ 
is given in Ref.~\cite{LOS}  
while the entanglement assisted capacity $C_E$ 
and a lower bound for $Q$ which is supposed to be thigh are given in Ref.~\cite{WERNER}.
Unfortunately, since an average input photon number constraint can not
prevent the average message of ${\cal E}_\eta$ from being supported on Fock states
with more than one photon (apart from the trivial case of zero average photon number),
the results obtained in \cite{LOS,WERNER}
provide only trivial upper bounds for the corresponding
capacities of ${\cal D}_\eta$. For instance consider the entanglement assisted
case where the capacities of both the channels can be computed. 
In the input Hilbert space we are considering here,
the average
photon number of the transmitted message is provided by the average population 
associated with the one photon Fock state. A fair comparison between the capacities of 
${\cal D}_\eta$ and ${\cal E}_\eta$  can be hence obtained by taking the value of $C_E$ 
associated with a lossy Bosonic ${\cal E}_\eta$ channel where the average
input photon number is given by the  population $p$ of Eq.~(\ref{finito2}) 
which maximizes the
entanglement-assisted capacity of ${\cal D}_\eta$. According to \cite{WERNER} the
capacity of ${\cal E}_\eta$  is then given by $C_E \equiv g(p) + 
g(\eta p) - g((1-\eta)p)$, where $g(x)=(x+1)\log_2(x+1)-x
\log_2 x$. A simple numerical analysis can be used to verify that this 
quantity is always bigger than the corresponding value~(\ref{finito2})
of ${\cal D}_\eta$.

The channels ${\cal D}_\eta$ and ${\cal E}_\eta$ 
share many common features. In particular
${\cal E}_\eta$ obeys to the same composition rule of ${\cal D}_\eta$
given in Eq.~(\ref{lossy3}) and it is degradable, since for
any input $\rho$ one has
\begin{eqnarray}
\tilde{{\cal E}}_\eta (\rho) = 
P S \; {{\cal E}}_{1-\eta} (\rho) \; S P
\label{lossy2APPENDIX}\;,
\end{eqnarray}
where now $P=e^{i\pi b^\dag b}$ and $S$ is the swap operator
which transforms $a$ in $b$ and vice versa.
This relation  was used in \cite{MINIMUM} without explicitly
proving it, and 
applies to all Gaussian channels of 
the form~(\ref{lossy}) where the external Bosonic
mode $b$ is prepared in a circularly symmetric
input. For the sake of completeness, here we 
give an explicit derivation of Eq.~(\ref{lossy2APPENDIX}) in 
the case of the purely lossy Bosonic channel.

Consider a generic input state $\rho$ of the Bosonic channel,
with characteristic 
function $\Gamma(\mu)\equiv \mbox{Tr}_a [ \rho D_a(\mu)]$, 
i.e.
\begin{eqnarray}
\rho \equiv \int \frac{d^2 \mu}{\pi} \; 
 \Gamma(\mu) \;  D_a(-\mu)
\label{car}\;,
\end{eqnarray}
where $D_a(\mu)\equiv \exp[\mu a^\dag - \mu^* a]$ 
is the displacement operator of the input mode $a$, 
\cite{WALLS}.
As shown in \cite{MINIMUM}, the channel ${\cal E}_\eta$ 
transforms $\rho$ into
\begin{eqnarray}
{\cal E}_\eta (\rho) = \int \frac{d^2 \mu}{\pi} \; 
 \Gamma^{\prime}(\mu) \;  D_a(-\mu)
\label{carstato}\;,
\end{eqnarray}
with
\begin{eqnarray}
\Gamma^{\prime} (\mu) \equiv \Gamma(\sqrt{\eta}\; \mu) \;
e^{-(1-\eta)|\mu|^2/2}
\label{car1}\;.
\end{eqnarray}
Analogously it is possible to verify that the complementary
map $\tilde{\cal E}_\eta$ (defined as in 
Eq.~(\ref{lossy1}) by replacing the partial trace over $b$ with
the partial trace over $a$ in the Eq.~(\ref{lossyAPPENDIX}))
produces the following transformation,
\begin{eqnarray}
\tilde{\cal E}_\eta (\rho) = \int \frac{d^2 \mu}{\pi} \; 
\tilde{\Gamma}^{\prime}(\mu) \;  D_b(-\mu)
\label{carstato1}\;,
\end{eqnarray}
where $D_b(\mu)$ is the displacement operator of the
mode $b$ and where
\begin{eqnarray}
\tilde{\Gamma}^{\prime}(\mu) 
\equiv \Gamma(-\sqrt{1-\eta}\; \mu) \;
e^{-\eta|\mu|^2/2}
\label{car2}\;.
\end{eqnarray}
Suppose now $\eta\geqslant 0.5$ and 
apply the lossy map 
${\cal E}_{1-\eta/\eta}$ to the state of 
Eq.~(\ref{carstato}): according to the composition rule
(\ref{lossy3}) this will transform its
symmetric characteristic function to
\begin{eqnarray}
{\Gamma}^{\prime\prime}(\mu) 
\equiv \Gamma( \sqrt{1-\eta}\; \mu) \;
e^{-\eta|\mu|^2/2} = \tilde{\Gamma}^{\prime}(-\mu)
\label{car2poi}\;,
\end{eqnarray}
producing the state
\begin{eqnarray}
{\cal E}_{\frac{1-\eta}{\eta}}({\cal E}_\eta (\rho)) 
= \int \frac{d^2 \mu}{\pi} \; 
\tilde{\Gamma}^{\prime}(\mu) \;  D_a(\mu)
\label{carstato2}\;.
\end{eqnarray}
This proves the identity~(\ref{lossy2APPENDIX}) since
under the unitary transformation
$P S$ the annihilation operator $a$ is transformed
into $-b$ and the state~(\ref{carstato2}) becomes
equal to the output~(\ref{carstato1}) 
of the composite map.

\section{Some useful relations}\label{s:app2}
In this appendix we provides some relations used
in Sec.~\ref{s:cap} to derive the capacities of
the qubit amplitude damping channel.

\subsection{Entropy of exchange}\label{s:app21}
The output state 
$({\cal D}_{\eta} \otimes\openone_{anc})  (\Phi)$ 
associated with the purification $|\Phi\rangle$ of
Eq.~(\ref{puri}) can be expressed in the
computational basis
$\{ | 0 0 \rangle, 
| 0 1  \rangle,
|1 0
\rangle,|1 1  \rangle \}$
of ${\cal H}_A\otimes  {\cal H}_{anc}$
(here $|0\rangle_{anc}\equiv |R_0\rangle$ while $|1\rangle_{anc}$ is
the component of $|R_1\rangle$ which is orthogonal to $|R_0\rangle$).
This gives the following $4\times 4$ matrix
\begin{widetext}
\begin{eqnarray}
\left( 
\begin{array}{cccc} 
1-p+(1-\eta)\,\frac{|\gamma|^2}{1-p}    & (1-\eta)\, \gamma\,
\frac{\sqrt{(1-p)p-|\gamma|^2}}{1-p} 
& \sqrt{\eta} \;\gamma^*& 
\sqrt{\eta}\, \sqrt{(1-p)p-|\gamma|^2} \\ \\
 (1-\eta)\, \gamma^*\,
\frac{\sqrt{(1-p)p-|\gamma|^2}}{1-p} 
& (1-\eta) \,(p -\frac{|\gamma|^2}{1-p})&0 & 0\\ \\
 \sqrt{\eta} \,\gamma & 0 & \eta\, \frac{|\gamma|^2}{1-p} &
\eta \, \gamma\,
\frac{\sqrt{(1-p)p-|\gamma|^2}}{1-p}  \\ \\
\sqrt{\eta}\, \sqrt{(1-p)p-|\gamma|^2}
& 0 & 
\eta \, \gamma^*\,
\frac{\sqrt{(1-p)p-|\gamma|^2}}{1-p} 
& \eta  \;
(p-\frac{|\gamma|^2}{1-p}) 
\end{array}
\right)\;.
\label{rhoout1}
\end{eqnarray}
\end{widetext}
The
matrix~(\ref{rhoout1}) has 
eigenvalues $0$ (two times degenerate) and 
\begin{eqnarray}
\Lambda_{\pm} &\equiv&
\left(1 \pm \sqrt{(1- 2(1-\eta) p)^2
+4(1-\eta) |\gamma|^2} \right)/2 \nonumber\\
&\equiv& \lambda_\pm(1-\eta)
\label{eigen1}\;,
\end{eqnarray}
with $\lambda_\pm$ the eigenvalues of ${\cal D}_\eta(\rho)$
given in Eq.~(\ref{eigen}).
The exchange entropy associated with
 $\rho$ is hence given by
\begin{eqnarray}
S(({\cal D}_{\eta} \otimes\openone_{anc})  (\Phi)) 
&=&-\Lambda_+ \ln \Lambda_+ -\Lambda_- \ln 
\Lambda_-\nonumber \\
&\equiv&
H_2 (\Lambda_+)\;,
\label{exentropy}
\end{eqnarray}
where, for $x\in[0,1]$,
\begin{eqnarray}
H_2(x) \equiv -x \log_2 x -(1-x) \log_2 (1-x)
\label{bin}
\end{eqnarray}
is the binary entropy \cite{COVER}.
Some useful relations of $H_2(z)$ are the following:
\begin{itemize}
\item[{\bf 1)}] The function 
$H_2(z)$ is decreasing with respect to the variable
$|1/2 + z|$. This property is a consequence of the
fact that
the entropy associated with a binary stochastic
 variable is maximum  when the probabilities 
associated with  different outcomes are equal.
\item[{\bf 2)}] The function $H_2((1+\sqrt{1- z^2})/2)$
is convex with respect to $z$. This property can
be easily verified and is related with the convexity
of the entanglement of formation with respect to concurrence
in qubit systems~\cite{WOT}. 
\end{itemize}
\subsection{Dependence on $\gamma$ of the coherent
information}\label{s:app22}
As discussed in the text, the quantum capacity $Q$ of
${\cal D}_{\eta}$ is obtained by maximizing the function
$J(p,|\gamma|^2)$ of Eq.~(\ref{COHERENTINFO})
over all values of
$p\in[0,1]$ 
and $|\gamma|^2\in[0,(1-p)p]$.
For $\eta\geqslant1/2$, this expression
is decreasing in $|\gamma|^2$, i.e. for any $|\beta|\in[0,1]$
it achieves the maximum value for $\gamma=0$. 
The proof of this result is quite tedious, but can
be obtained analytically by studying the 
partial derivative of~(\ref{COHERENTINFO}) with
respect to the parameter $|\gamma|^2$.
We skip all the details of this analysis which is
not of a fundamental interest,  
and simply observe that
the problem can be reduced to
studying the properties of the function
\begin{eqnarray}
f_y(x) \equiv \left(\frac{1 - \sqrt{(1- 2x)^2
+4 x y }}{1 + \sqrt{(1- 2x)^2
+4 x y }} \right)^{{x}/
{\sqrt{(1- 2 x)^2
+4xy}}}
\label{coherent3}
\end{eqnarray}
on the domain $x\in[0,1-y]$, for any $y\in [0,1]$.
One can then verify that $f_y(x)$ is decreasing in $x$:
this guarantees that 
$J(p,|\gamma|^2)$ is monotonically 
decreasing in $|\gamma|^2$, yielding the thesis.


\begin{thebibliography}{99}
\bibitem{NOTABENE}
Any unitary transformation of the output signal can be 
in fact reversed by the receiver at the decoding stage.
In the case of Eq.~(\ref{rhodibeta1}) this can be
accomplished by applying to $B$ a unitary operator  
which rotates   $|\phi_1^\prime\rangle_B$ of Eq.~(\ref{PHIPRIME1})
into $|\phi_1\rangle_B\equiv \sum_{j=N-k+1}^N c_j|{ j}\rangle$  
while leaving  $|\Downarrow \; \rangle_B$ unchanged.
\bibitem{NOTAPARALLEL} 
In the context of memoryless channel 
the parameter  $n$ is often presented as 
the number of ``successive'' uses of the
channel  \cite{SHOR,CHUANG}.  However, in the spin chain  
model we are considering here this definition is misleading since
repetitive uses of the same chain introduce memory effects 
that are not take into account by Eq.~(\ref{mappa}) 
(see also Sec.~\ref{s:uno}).
For this reason we refer to $n$ as the number ``parallel'' uses of the
channel to stress the fact that the 
sender is operating at the same time on $n$
identical ``copies'' of the original spin chain.
\bibitem{ARGO}
For $\eta_1\leqslant \eta_2$ Eq.~(\ref{degrado2}) can be reversed
to show that by concatenating the map  $\tilde{\cal T}_{\eta_1,\eta_2}$ 
with an amplitude damping channel of transmissivity $\eta_1/\eta_2$ on can simulate
any output generated from ${\cal T}_{\eta_1,\eta_2}$, i.e.
${\cal T}_{\eta_1,\eta_2} (\rho) =  S\; 
{\cal D}_{\eta_1/\eta_2} (\tilde{\cal T}_{\eta_1,\eta_2} (\rho))\;S $.
This means that if by contradiction ${\cal T}_{\eta_1,\eta_2}$ would have a no null 
quantum capacity then the sender would be able to transmit
two copies of the same quantum information at the output of the two 
channels ${\cal T}_{\eta_1,\eta_2}$ and $\tilde{\cal T}_{\eta_1,\eta_2}$.
\end{thebibliography}
\end{document}